 \documentclass[final,5p,times,twocolumn]{elsarticle}

\usepackage[dvipsnames]{xcolor}
\usepackage{amssymb}
\usepackage{lipsum}
\usepackage{amsthm}
\usepackage{amsmath} 
\usepackage{slashed} 
\usepackage{cuted} 
\usepackage{color}

\usepackage[colorlinks=true]{hyperref}
\makeatletter
\AtBeginDocument{%
  \@ifpackageloaded{hyperref}{%
    \def\@linkcolor{magenta}%
    \def\@anchorcolor{blue}%
    \def\@citecolor{cyan}%
    \def\@filecolor{cyan}%
    \def\@urlcolor{cyan}%
    \def\@menucolor{blue}%
    \def\@pagecolor{blue}%
  }{}
}
\makeatother

\newcommand{\ba}{\begin{array}}
\newcommand{\ea}{\end{array}}
\newcommand{\ipc}[1]{\textcolor{Maroon}{#1}}

\journal{Physics Letters B}

\begin{document}

\begin{frontmatter}



\title{On modified Hawking radiation under quantum gravity effects in the fermion sector}


\author[first]{I. P-Castro}
\ead{ivan.perez.c@cinvestav.mx}
\author[first]{H. Aguilera-Trujillo}
\ead{hector.aguilera@cinvestav.mx}
\author[first]{H. García-Compeán}
\ead{hugo.compean@cinvestav.mx}
\author[first]{A. Pérez-Lorenzana}
\ead{abdel.perez@cinvestav.mx}
\affiliation[first]{organization={Departamento de Física, Centro de Investigación y de Estudios Avanzados del I.P.N.},
            addressline={Apartado Postal 14-740}, 
            city={Ciudad de México},
            postcode={07000}, 
            state={},
            country={México}}

\begin{abstract}
In this letter, after briefly discussing Hawking radiation (HR) as quantum tunneling, we emphasize that special care must be taken when computing temperature corrections in the presence of Lorentz invariance violation (LIV), particularly by ensuring consistency between the modified dispersion relation and the gamma matrix structure used to describe fermions that are tunneling. We present a consistent implementation of such corrections in static, spherically symmetric black hole (BH) spacetimes. As the conserved Killing energy $\omega = p_\mu \xi^\mu$ in a stationary, asymptotically flat spacetime corresponds to the particle's energy measured by an observer at infinity, one might expect LIV terms that increase (decrease) the energy of a free particle, as defined in the local spacetime frame, to lead to a corresponding increase (decrease) in the Hawking temperature. We discuss the conditions under which this expectation holds.
We also determined the changes in Bekenstein-Hawking entropy for Schwarzschild-like BHs, including corrections from LIV scenarios.
\end{abstract}



\begin{keyword}
Quantum tunneling \sep Hawking radiation \sep Lorentz invariance violation \sep Quantum gravity effects



\end{keyword}

\end{frontmatter}




\section{Introduction}
\label{introduction}
Hawking’s works in 1974-75~\cite{Hawking1974black, Hawking1975particle} demonstrated that BHs are not really black because they emit thermal radiation with a temperature proportional to their surface gravity $\kappa$, establishing a deep connection between quantum mechanics, general relativity, and thermodynamics~\cite{Bardeen1973four}. This phenomenon, known as the {\it Hawking effect}, arises from the observer-dependent nature of particle definitions in curved spacetime, particularly near the event horizon, where infalling modes become causally disconnected from the exterior.

A particularly intuitive explanation of HR involves particles' quantum tunneling across a BH's event horizon. In this picture, particle–antiparticle pairs are created in the vicinity of the horizon, with one particle escaping along a classically forbidden path characterized by a complex action. Because the dynamics are confined to the near-horizon region and proceed primarily in the radial direction, the system effectively reduces to a two-dimensional chiral theory in the $t-r$ sector of the metric, where only the $s$-wave mode is relevant. This dimensional reduction is key to the emergence of gauge and gravitational (diffeomorphism) anomalies, which ultimately give rise to the Hawking flux~\cite{Robinson2005relationship,Iso2006hawking, Banerjee2009connecting}. In this approach, two main methods have been developed to compute the imaginary part of the complex action: the radial null geodesic method~\cite{Parikh2000hawking, Parikh2004secret} and the Hamilton-Jacobi (HJ) approach~\cite{Angheben2005hawking}~\footnote{That is an extension of the complex path analysis of Refs.~\cite{Srinivasan1999particle,Shankaranarayanan2001method}.}.

In the broader context of quantum gravity (QG), theories like loop quantum gravity suggest that Lorentz invariance might be broken at high energies, leading to modified dispersion relations (MDRs) suppressed by the Planck scale~\cite{Alfaro2000quantum,Alfaro2002quantum}. This is relevant for enriching BH physics and probing observable QG effects~\cite{Amelino2002relativity, Amelino2005search, Jacobson2003strong}. These MDRs affect fundamental quantities such as the BH temperature and Bekenstein-Hawking entropy. Prior studies have explored the impact of MDRs on fermionic tunneling from stationary BHs, revealing non-trivial corrections to the process~\cite{Yang2016lorentz, Feng2019modified, Kamali2016radiation}. 

The art of these corrections involves starting with a modified dispersion relation (MDR) in flat spacetime that increases or decreases the energy of the free particle, identifying the corresponding equation of motion, and generalizing it using the covariant derivative. It is intuitive to expect that this behavior persists in curved spacetime, with particles of lower (or higher) energy being emitted. However, this is not always found in the literature~\cite{Li2017application, Li2021lorentz}. Motivated by these observations, we investigate the origin of these discrepancies and demonstrate that this effect emerges whenever the MDR is uniquely determined.

On the other hand, in the context of string theory, effective field theories that describe LIV can be constructed as extensions of the Standard Model of particle physics, collectively known as the Standard-Model Extension (SME).~\cite{Colladay1997cpt, Colladay1998lorentz}. In the simplest non-degenerate case, each spin projection of particles and antiparticles is associated with a distinct energy, leading to different effective Hawking temperatures for fermions and antifermions, suggesting a possible mechanism for explaining the baryon asymmetry of the Universe.
\section{Quantum tunneling approach for Hawking radiation}
\label{QT-HR}
According to the formulation by Hawking and Gibbons~\cite{Gibbons1977action}, the probability of particle emission, $P[out]$, from the past horizon is not equal to the probability of absorption into the future horizon, $P[in]$. The ratio between these probabilities is~\cite{ Srinivasan1999particle}
    \begin{equation}
        \Gamma = \frac{P[out]}{P[in]} = e^{-\omega/T_H} \, ,
    \end{equation}
where $\omega$ is the conserved Killing energy, corresponding to the particle's energy measured by an observer at infinity in an asymptotically flat, stationary spacetime, and $T_H$ denotes the Hawking temperature. The above relation exhibits the characteristic thermal Boltzmann factor of HR, analogous to that associated with blackbody radiation at temperature $T_H$.

In the tunneling picture, HR originates from particle-antiparticle pair creation in the vicinity of the event horizon. Pair creation strictly inside or outside the horizon leads to apparent causal and coordinate-time paradoxes. These are resolved in the Parikh--Wilczek framework~\cite{Parikh2000hawking, Parikh2004secret} by accounting for the dynamical shrinkage of the horizon during emission. A detailed discussion of the consistency of the tunneling approach is given in Ref.~\cite{Vanzo2011tunneling}. Within this framework, the tunneling probability is determined by the imaginary part of the semiclassical action $I$, which yields the Boltzmann factor associated with the Hawking temperature~\cite{Parikh2000hawking},
    \begin{equation}
        \Gamma = \exp\{ -2Im I\} = \exp\{-\omega/T_H\}\, .
    \end{equation}
In the reformulation of quantum tunneling presented in~\cite{Banerjee2008quantum, Majhi2009fermion}, $P[out]$ denotes the probability for the particle to tunnel through the classically forbidden trajectory, while $P[in]$ corresponds to the probability that the ingoing partner is absorbed by the BH. A well-known subtlety of this formalism is the so-called ``factor-of-two problem''~\cite{Akhmedov2006hawking}, which has been addressed through several approaches, including accounting for the imaginary contribution of the temporal part of the action~\cite{Akhmedov2008subtleties}. However, the simplest solution in the HJ method involves introducing an additional constant to the HJ ansatz for the classical action, denoted by $K$, to compensate for the ``imaginary contribution of time'' and ensure $P[in] = 1$~\cite{Mitra2007hawking}. This is necessary only when working in coordinate systems that do not cover the event horizon, that is, when the coordinates exhibit singularities at the event horizon. 

The semiclassical HJ approach is the simplest applicable method for introducing QG corrections. We would like to remark that early studies suggested that an expansion of the semiclassical action in powers of $\hbar$~\cite{Banerjee2008quantum, Majhi2009fermion} could give rise to higher-order quantum corrections to the Hawking spectrum. However, it has subsequently been shown that such corrections are not physical when the energy is consistently defined within the HJ formalism as $\omega = - \partial_t I$~\cite{Wang2010hawking, Chatterjee2009hawking, Yale2011there}. In the following, we apply the HJ method to fermions.
\section{From flat spacetime to curved spacetime} \label{DR-to-Cspacetime}
We will use $a \in \{0,1,2,3\}$ to denote local indices, and $\mu \in \{t, r, \theta, \varphi\}$ to denote curved spacetime indices, taking the West Coast signature convention $(+,-,-,-)$.

The Lorentz invariance of the Dirac equation, $(i\tilde{\gamma}^a \partial_a - m) \psi = 0$, and of the Dirac-like equation for gravitinos, $(i\tilde{\gamma}^a \partial_a - m) \psi_b = 0$, leads to the usual dispersion relation (DR) for a free massive particle, $p_0^2 = m^2 + {\bf p}^2$. For the latter case, the constraint $\tilde{\gamma}^b \psi_b =0$, removes non-physical states to describe spin-$\frac{3}{2}$ particles.
 
As discussed in the Introduction, QG effects are commonly explored through modifications of the DR. One such modification is the generalization known as \emph{deformed relativity} (dR), which is given by
    \begin{equation}
        p_0^2 = m^2 + {\bf p}^2 - \eta_{\pm} (L p_0)^n {\bf p}^2 \, , 
        \label{MDR-DR}
    \end{equation}
Here, $\eta_{\pm} = \pm 1$, and $L = \frac{\ell_p}{\xi_\alpha}$, where $\ell_p$ is the Planck length and $\xi_\alpha$ is a dimensionless parameter that depends on the particle and the model. For $n = 2$ and $\eta_{+}$, the modified Dirac equation is
    \begin{equation}
        (i \tilde{\gamma}^a \partial_a - m -L\tilde{\gamma}^0 \partial_0 \tilde{\gamma}^l \partial_l) \psi =0 \, . 
        \label{LIV-DR-fermions}
    \end{equation}
This equation is equivalent to the one presented in~\cite{Kruglov2012modified}. Notice that, as the author indicates, he uses a Euclidean metric. We can obtain the MDR by considering solutions of the form $\psi(x) = \tilde{\psi}(\mathbf{p}) e^{-i p_c x^c}$ for fermions. We then compute the determinant of the resulting operator acting on $\tilde{\psi}({\bf p})$ and set it to zero. Similarly for gravitinos
    \begin{equation}
        (i \tilde{\gamma}^a \partial_a - m - L\tilde{\gamma}^0 \partial_0 \tilde{\gamma}^l \partial_l )\psi_b =0 \, .
        \label{LIV-DR-gravitinos}
    \end{equation}
The modified Dirac equations~\eqref{LIV-DR-fermions} and \eqref{LIV-DR-gravitinos} reproduce the MDR given in Eq.~\eqref{MDR-DR}. A similar MDR can be obtained by replacing $L \to -L$. Then, the MDR is not uniquely determined.

On the other hand, for the SME at order $k=0$~\cite{Colladay1997cpt}, the relevant contributions are
    \begin{equation}
        (i \tilde{\gamma}^a \partial_a - m -\slashed{a} - \tilde{\gamma}_5 \slashed{b})\psi = 0 \, ,
        \label{SME-0-fermions}
    \end{equation}
with background vectors $a_c$ and $b_c$. In the most general case, the MDR is quite complex, but we can extract an interesting case by setting $b_c = (b_0, {\bf 0})$, valid if $b^c$ is a timelike vector, which gives a non-degenerate energy spectrum for free particles~\cite{Lehnert2004dirac},
    \begin{align}
        E_u^{(\alpha)} &= [m^2 + \left( |{\bf p} - {\bf a}| + (-1)^\alpha b_0 \right)^2]^{1/2} + a_0 \, , \label{f-E} \\ 
        E_v^{(\alpha)} &= [m^2 + \left( |{\bf p} + {\bf a}| + (-1)^\alpha b_0 \right)^2]^{1/2} - a_0 \, ,
        \label{anti-f-E}
    \end{align}
where $\alpha = 1,2$, corresponding to spin-up ($\alpha = 1$) and spin-down ($\alpha = 2$), for example. Here, we have used the $C$-symmetry properties: for fermions, the four-vector are $\{a_c, b_c\}$, and for antifermions, $\{-a_c, b_c\}$. At leading order, the previous expressions become 
    \begin{equation}
        E_u^{(1,2)} = \sqrt{m^2 + {\bf p}^2} + a_0 - \frac{{\bf p} \cdot {\bf a} \pm b_0|{\bf p}|}{\sqrt{m^2 + {\bf p}^2}} \, ,
        \label{SME-first}
    \end{equation}
provided the upper
and lower signs in Eq.~\eqref{SME-first} are identified with the labels $\alpha =1$ and $\alpha=2$, respectively. As we explain before $a_c \to -a_c$ describes $E_v^{(1,2)}$. A similar framework has recently been developed to describe spin-$\frac{3}{2}$ particles~\cite{P2025cpt}, such as gravitinos. In this case, one can construct operators analogous to those described above
    \begin{equation}
        (i \tilde{\gamma}^a \partial_a - m -\slashed{a} - \tilde{\gamma}_5 \slashed{b})\psi_b = 0 \, .
        \label{SME-0-gravitinos}
    \end{equation}
Again from $C$-symmetry, an energy spectrum of the form \eqref{f-E} and \eqref{anti-f-E} can be derived from the solution $\psi_b(x) = \tilde{\psi}_b(\mathbf{k}) e^{-ik_\mu x^\mu}$. While a precise interpretation of the labels $ (\alpha)$ is still lacking, this is not relevant for our purposes; what matters is that the energies can be associated with the distinction between gravitinos ($ a_c$) and antigravitinos ($-a_c$). We emphasize that our analysis assumes Dirac gravitinos, in contrast to the minimal supergravity framework, which requires Majorana gravitinos~\cite{Freedman2012supergravity}.

To formulate a theory of fermions on a curved spacetime, we cannot work directly in general coordinates due to the absence of spinor representations of $GL(4,\mathbb{R})$. Instead, we introduce a local orthonormal frame, the tetrad formalism, which relates the spacetime gamma matrices to the flat-space ones via $\gamma^\mu = e^\mu_a \tilde{\gamma}^a$. This reduces the structure group of the frame bundle from $GL(4,\mathbb{R})$ to $SO(1,3)$. Spinor fields are then realized as sections of vector bundles associated to a principal $\mathrm{Spin}(1,3)$-bundle through a spinor representation of $\mathrm{Spin}(1,3)$.

In phenomenological approaches based on MDRs, a preferred frame must be specified to covariantly define the energy and spatial momentum entering the MDR~\cite{Tao2017black, Wang2016black}. Then, the local energy and momentum in Eq.~\eqref{MDR-DR} are measured with respect to this frame, which is defined by a unit timelike vector
$u^\mu$ ($u_\mu u^\mu = 1$) tangent to the worldlines of the preferred observers. Accordingly, the local energy is $E_{\rm loc}=p_\mu u^\mu$, and the magnitude of the spatial momentum is given by $p^2 = E_{\rm loc}^2 - p_\mu p^\mu$, where $p^\mu$ is the four-momentum.

We generalize the Dirac equations to arbitrary curved spacetime using generalized gamma matrices and a covariant derivative. In dR (with $\hslash \neq 1$)
    \begin{align}
        \left( i \gamma^\mu \nabla_\mu - \frac{m}{\hslash} - \hslash L \gamma^t \nabla_ t \gamma^j \nabla_j \right) \psi &= 0 \, , \label{LIV-DR-12}\\
        \left( i \gamma^\mu \nabla_\mu - \frac{m}{\hslash} - \hslash L \gamma^t \nabla_ t \gamma^j \nabla_j \right) \Psi_\nu &= 0 \label{LIV-DR-32}\, . 
    \end{align} 
In contrast, within the SME, Lorentz violation is encoded directly through background coefficient fields, such as $a_\mu^f = e_\mu^c a_c^f$~\cite{Kostelecky2004gravity, Kostelecky2021backgrounds}; therefore, the tetrad $e_\mu^c$ is fixed by a preferred frame. We discuss the challenges associated with constructing such background vectors later.

We obtain the following equations of motion:
    \begin{align} 
        \left(i \gamma^\mu \nabla_\mu - \frac{m}{\hslash} -\frac{a_\mu^f}{\hslash} \gamma^\mu -  \frac{b_\mu^f}{\hslash} \gamma_5 \gamma^\mu \right) \psi &= 0 \,.
        \label{LIV-SME-12} \\
        \left(i \gamma^\mu \nabla_\mu - \frac{m}{\hslash} -\frac{a_\mu^f}{\hslash} \gamma^\mu -  \frac{b_\mu^f}{\hslash} \gamma_5 \gamma^\mu \right) \Psi_\nu &= 0 \,.
        \label{LIV-SME-32}
    \end{align}
Here, $\nabla_\mu = \partial_\mu - i\frac{q}{\hslash} A_\mu + \Omega_\mu$ for spin-$\frac{1}{2}$ particles, and for massive gravitinos $\nabla_\mu \Psi_\nu = \partial_\mu \Psi_\nu - \Gamma_{\mu\nu}^\lambda \Psi_\lambda$ with the condition $\gamma^\mu \Psi_\mu =0$, where $\Omega_\mu = \frac{1}{4} \omega_{\mu ab} \tilde{\gamma}^{ab}$ is the spin connection, $f$ is a label that refers to type of fermions. 
\subsection{Static, spherically symmetric black hole and generalized gamma matrices} \label{QT-exp}
Consider a static, spherically symmetric spacetime in standard Boyer-Lindquist (BL) coordinates $(t, r, \theta, \varphi)$,
    \begin{equation}
        ds^2 = f(r) \, dt^2 - \frac{1}{g(r)} \, dr^2 - r^2 d\Omega^2 \, ,
        \label{RN-like metric}
    \end{equation}
where $d\Omega^2 = d\theta^2 + \sin^2\theta \, d\varphi^2$. The vacuum solution (i.e., no matter fields, $T_{\mu\nu} = 0$) of the Einstein-Maxwell field equations reduces to $R_{\mu\nu} = 0$, from which one can show that $f(r) = g(r)$ up to an overall constant that can be absorbed by rescaling the time coordinate. In the presence of a cosmological constant, $\Lambda$, the ``vacuum'' Einstein equations become $R_{\mu\nu} = \Lambda g_{\mu\nu}$. Due to the static and spherically symmetric nature of the metric, the off-diagonal components of the Ricci tensor vanish, and the structure of the remaining equations still enforces $f(r) = g(r)$, again up to a rescaling of time. The most general metric for a charged BH with a cosmological constant is given by\footnote{In Planck units.}
    \begin{equation}
        f(r) = 1 - \frac{2M}{r} + \frac{Q^2}{r^2} - \frac{\Lambda}{3} r^2 \, ,
    \end{equation}
where $M$ is the BH mass, $Q$ is the electric charge, and $\Lambda$ is the cosmological constant, which modifies the global structure of spacetime. When $\Lambda = 0$, the metric~\eqref{RN-like metric} reduces to the Reissner-Nordström BH (RN-BH). If $\Lambda > 0$, the solution corresponds to an asymptotically \textit{de Sitter} (dS) space, which implies a universe with accelerated expansion, and there is a cosmological horizon (CH), beyond which distances expand so rapidly that nothing can escape. \ipc{For} $\Lambda <0$, we obtain an asymptotically \textit{Anti-de Sitter} (AdS) space, characterized by negative spatial curvature. In contrast to dS space, pure AdS ($M=0$ and $Q=0$) does not exhibit a CH, as signals can, in principle, reach the conformal boundary.

To derive a deformed HJ equation incorporating the MDR, one must first specify the preferred frame in the black hole spacetime. A natural choice is the static frame associated with observers hovering at a fixed radial coordinate. In de Sitter spacetimes, however, this choice is valid only within the static patch, since the static frame does not extend beyond the cosmological horizon. In this frame, the four-velocity is
    \begin{equation}
        u^\mu = \left(\sqrt{g^{tt}}, {\bf 0}\right) \, . \label{Static frame}
    \end{equation}
If $I$ is a solution of the deformed HJ equation, the four-momentum is related to the classical action by $p_\mu = - \partial_\mu I$~\cite{Tao2017black}.

In BL coordinates, using $g^{\mu\nu} = e^\mu_a e^\nu_b \eta^{ab}$ in the tetrad formalism, we can construct the generalized gamma matrices as
    \begin{equation}
        \gamma^t = \frac{1}{\sqrt{f(r)}} \tilde{\gamma}^0 \, , \, \gamma^r = \sqrt{f(r)} \tilde{\gamma}^3 \, , \, \gamma^\theta = \frac{1}{r} \tilde{\gamma}^1 \, , \, \gamma^\varphi = \frac{1}{r\sin\theta} \tilde{\gamma}^2 \, .
        \label{Gamma-Sing}
    \end{equation}
This canonical choice describes the static frame.

The metric~\eqref{RN-like metric} has a singularity due to the choice of coordinates, so we can choose generalized {\it Painlevé--Gullstrand} (PG) coordinates through the transformation:
    \begin{equation}
        dt \to dt_p = dt - \sqrt{\frac{1-f(r)}{f^2(r)}} dr \, .
    \end{equation}
So the Schwarzschild-like metric becomes\footnote{Here we use $t$ instead of $t_p$; we use the latter only when there is a substantial distinction.}
    \begin{equation}
        ds^2 = f(r) dt^2 - 2 \sqrt{1-f(r)} dt dr - dr^2 - r^2 d\Omega^2 \, .
    \end{equation}
And we can choose the freely falling observers, which are tangent to the infalling worldlines, with $v(r) = - \sqrt{1-f(r)}$~\cite{Wang2016black}; we have
    \begin{equation}
        u^\mu = \left(1, v(r), 0,0 \right)  \, .
    \end{equation}
Then, using PG coordinates, the tetrad that describes a freely falling observer is
    \begin{equation}
        \gamma^t = \tilde{\gamma}^0 \quad \text{and} \quad \gamma^r = - \sqrt{1-f(r)} \tilde{\gamma}^0 + \tilde{\gamma}^3 \, .
        \label{Gamma-P}
    \end{equation}
While the static frame is described by
    \begin{equation}
        \gamma^t = \frac{1}{\sqrt{f(r)}} \left( \tilde{\gamma}^0 + \sqrt{1-f(r)} \tilde{\gamma}^3 \right) \quad \text{and} \quad \gamma^r = \sqrt{f(r)} \tilde{\gamma}^3 \, ,
        \label{Gamma-GP}
    \end{equation}
while $\gamma^\theta$ and $\gamma^\varphi$ remained equal to the one given in Eq. ~\eqref{Gamma-Sing}.

In the chiral representation of the gamma matrices, adopting the West Coast metric signature, we have
    \begin{equation}
        \tilde{\gamma}^a = \left(\ba{cc} 0 & \sigma^a \\
        \bar{\sigma}^a & 0 \ea\right) \, ,
        \label{Gamma-chiral}
    \end{equation}
where $\sigma^a =({\bf 1}, \sigma^l)$, $\bar{\sigma}^a =({\bf 1}, -\sigma^l)$, and $\sigma^l$ with $l\in\{1,2,3\}$ are the standard Pauli matrices. 

In the following, we adopt this representation irrespective of the frame, as it allows us to use the usual ansatz~\cite{Kerner2008fermions, Yale2009gravitinos, Chen2013fermion, Liu2014fermions}. Other representations are physically equivalent and related by a unitary transformation, $U$; $\tilde{\gamma}_a = U \tilde{\gamma}_a^\prime U^\dagger$, $\psi=U\psi^\prime$, $\Psi_\nu = U\Psi_\nu^\prime$.
\section{Fermion tunneling in a static and spherically symmetric black hole}
\label{QT}
Near the event horizon (or cosmological horizon), the propagation of high-frequency modes can be described within the geometric-optics approximation. We therefore employ the Wentzel–Kramers–Brillouin (WKB) approximation to analyze the semiclassical system. We adopt the following WKB ansatz, generalizing the spinor solution in the chiral basis
(see Eq.~\eqref{Chiral-spinors-radial} of the Supplementary Material):
    \begin{equation}
        \psi_{\uparrow} = \left[\ba{c} A \\
        0 \\
        B  \\
        0 \ea\right] \exp\left\{ \frac{i}{\hslash} I_{\uparrow} \right\} \, , \quad
        \psi_{\downarrow} = \left[\ba{c} 0 \\C \\
        0 \\
        D  \ea\right] \exp\left\{ \frac{i}{\hslash} I_{\downarrow} \right\} \, .  \label{12Fermion-ansatz}
    \end{equation}
where $\uparrow$ ($\downarrow$) corresponds to spin-up (spin-down) projection, and $I_{\uparrow /\downarrow}$ is the semiclassical action, and we omitted to write the dependence of coordinates $(t,\theta, \varphi,r)$ of $\psi_{\uparrow/\downarrow}$, $A, B, C, D$ and $I_{\uparrow/\downarrow}$.

One might think that we can use similar reasoning to work with gravitinos, as in Eq.~\eqref{Gravitinos-FS} presented in the supplementary material; this would require constructing spin projection operators in curved spacetime, which would be complicated. However, this is not necessary, and we can work with the WKB ansatz \begin{equation}
        \Psi_\mu = \left[\ba{c} a_\mu \\
        b_\mu \\
        c_\mu \\
        d_\mu \ea\right] \exp\left\{ \frac{i}{\hslash} I \right\}  \, .
        \label{gravitino-ansatz}
    \end{equation}
with the constraint $\gamma^\mu \Psi_\mu =0$, and here $I$ is the semiclassical action. Again, we omitted the dependence on $(t,\theta, \varphi,r)$.

When a charged particle tunnels through the potential barrier, the self-interaction effects of the electromagnetic field on the emitted particles must be taken into account; then $A_t \neq 0$. However, in the PG coordinates, the electromagnetic potential is ill-defined at the horizon; for this reason, we use gauge invariance, $A_\mu \to A_\mu + \partial_\mu \chi$, to regularize the electromagnetic potential, $A_t(r) = -Q/r$. 

In the following analysis, we neglect the effects of energy and charge conservation in the BH–particle system when determining the modified Hawking temperature (see the Supplementary Material for technical details).
\subsection{Fermions tunneling under deformed relativity}
We can proceed to solve the modified Dirac equation~\eqref{LIV-SME-12} for the spin-up case.
From the ansatz~\eqref{12Fermion-ansatz}, we can reduce the motion equation in the limit $\hslash \to 0$. Using the regular coordinates~\eqref{Gamma-P} of a freely falling observer and the HJ ansatz
     \begin{equation}
        I_{\uparrow} = -\omega t + W(r) + J(\theta, \varphi) \, ,
        \label{HJ-spin-up}
    \end{equation} 
we obtain a non-separable system of equations in $t$ and $r$:
\begin{equation}
\begin{aligned}
    B &\left[ -(\omega + q A_t) + \left( 1 - \sqrt{1-f(r)}\right)(\partial_r W) \right] \\
    &=- A \left[ m - \left( \sqrt{1-f(r)} + 1 \right) (\omega + qA_t) L ( \partial_r W) \right] ,\\
    A &\left[ -(\omega + q A_t) - \left( 1 + \sqrt{1-f(r)} \right) (\partial_r W) \right] \\
    &= -B \left[ m -  \left( \sqrt{1-f(r)} - 1 \right) (\omega +qA_t) L (\partial_r W) \right] \, . 
\end{aligned} \label{Fermions-up-sys}
\end{equation}
While the angular part is described by $(\partial_\theta J) + \frac{i}{\sin\theta} (\partial_\varphi J) =0$. 

To obtain non-trivial solutions of~\eqref{Fermions-up-sys}, we solve the system by setting the determinant equal to zero, which yields:
    \begin{equation}
        \partial_r W_{\pm}(r) = \frac{B(r) \pm \sqrt{\Delta(r)}}{A(r)} \, ,
        \label{W-spin-up}
    \end{equation}
here $A(r) =  f(r) [ 1 - (\omega + qA_t)^2 L^2 ]$ , $
C(r) = m^2 - (\omega + qA_t)^2$ , $
B(r) =  \sqrt{1-f(r)} (\omega + q A_t)(1 + mL) $ , $\Delta(r) = [B(r)]^2 - A(r) C(r)$.
Near the horizon, we can approximate $f(r)\approx f^\prime(r_H)(r-r_H)$ and integrating around the simple pole, $r=r_H$, we obtain
    \begin{equation}
        W_{+} = \frac{2i\pi E (1 + mL)}{f^\prime(r_H) (1 - E^2 L^2)} \quad \text{and} \quad W_{-} = 0 \, . 
    \end{equation}
Therefore, the correct Hawking temperature is
    \begin{equation} 
        T_H = \frac{f^\prime(r_H)}{4\pi} \frac{(1 - E^2 L^2)}{(1 + mL)} \, ,
        \label{MHT-fermions}
    \end{equation}
where $E = \omega + q A_t(r_H)$, for non-charged fermions or $Q=0$; $E=\omega$. The same result is obtained using the representation for a static observer~\eqref{Gamma-GP} and/or considering the spin-down case.  
\subsection{Gravitinos tunneling under deformed relativity}
Now, we proceed to explore the same scenario, but now with gravitinos.
    
To analyze the classical limit ($\hbar \to 0$), we have used the ansatz~\eqref{gravitino-ansatz} to reduce the HJ equation; we focus explicitly on the static observer~\eqref{Gamma-GP} and the HJ ansatz takes the form
    \begin{equation}
        I = -\omega t + W(r) + J(\theta, \varphi) \, .
        \label{HJ-ansatz-gravitino}
    \end{equation}
We obtain in the $t-r$ regime for coefficients $a_{\nu}$ and $c_{\nu}$
\begin{equation}
\begin{aligned}
    c_{\nu} &\left[ -\frac{1}{\sqrt{f(r)}} \left(1 + \sqrt{1-f(r)} \right) \omega + \sqrt{f(r)} (\partial_r W) \right] \\
    &= - a_{\nu} \left[ m - \left( 1 + \sqrt{1-f(r)} \right) \omega L (\partial_r W) \right] \, ,\\
    a_{\nu} &\left[ - \frac{1}{\sqrt{f(r)}} \left(1 - \sqrt{1-f(r)}\right) \omega - \sqrt{f(r)} (\partial_r W)\right] \\ 
    &= - c_{\nu} \left[ m + \left( 1- \sqrt{1-f(r)} \right) \omega L( \partial_r W) \right]  \, .
\end{aligned} \label{Gravitino-sys}
\end{equation}
Notice that we obtain another equivalent system of equations by replacing $a_{\nu} \to d_{\nu}$ and $c_{\nu} \to b_{\nu}$, so in either case we arrive at the same set of solutions that are analogous to Eq.~\eqref{W-spin-up}.

Near the horizon, $f(r) \approx f^\prime(r_H) (r-r_H)$; integrating around the simple pole $r=r_H$ as in the previous example, we obtain the correct Hawking temperature for any component, in particular for physical components that describe gravitinos:
    \begin{equation} 
        T_H = \frac{f^\prime(r_H)}{4\pi} \frac{(1 - \omega^2 L^2)}{(1 + mL)} \, .
        \label{MHT-gravitinos-L}
    \end{equation}
Notice that the results in Eqs.~\eqref{MHT-fermions} and \eqref{MHT-gravitinos-L} have the same form regardless of whether the preferred frame is chosen to be that of a freely falling or a static observer. We discuss why the PG coordinates allow us to recover the correct spectra. Before doing so, however, we consider another example later.
\subsection{Fermions tunneling in a static spherical black hole under SME}
\label{QT-LIV-SME}
Consider the fermions described by Eq.~\eqref{LIV-SME-12} and Eq.~\eqref{LIV-DR-32}. As explained in Section~\ref{DR-to-Cspacetime}, we work under the condition $b_\mu^f = (b_t,{\bf 0})$ for simplicity, and to avoid interfering with the angular part, we choose the background vector $a_\mu^f = (a_t, 0, 0, a_r)$. Thus, in the PG coordinate for a freely falling observed~\eqref{Gamma-P}, we have
    \begin{equation}
        \gamma_5 = \frac{1}{r^2 \sin\theta} \left[ -\sqrt{1-f(r)} \left(\ba{cc} \sigma^3 & 0 \\
        0 & \sigma^3 \ea \right) + \left(\ba{cc} -{\bf 1} & 0 \\
        0 & {\bf 1} \ea\right)\right] \, .
        \label{gamma5-P}
    \end{equation}
In the following, we omit the explicit labels; however, it should be noted that the coefficients may differ for each fermion. And we consider that $a_t(r=r_H)$ has a finite value.
    
For spin-$\frac{1}{2}$, from the WKB ansatz \eqref{12Fermion-ansatz}, using the gamma matrices representation \eqref{Gamma-P}, with~\eqref{gamma5-P}, and the HJ ansatz \eqref{HJ-spin-up}, we obtain the following system of equations in $t-r$ regime
    \begin{equation}
    \begin{aligned}
        B &\bigg[ -(\omega + qA_t - a_t) + ( 1 - \sqrt{1-f(r)}) (\partial_r W + a_r) \\
        &\quad + \frac{b_t}{r^2 \sin\theta} (\sqrt{1-f(r)} -1) \bigg] + A m = 0 \, , \\
        A &\bigg[ -(\omega + qA_t - a_t) - ( 1 + \sqrt{1-f(r)}) (\partial_r W + a_r) \\
        &\quad - \frac{b_t}{r^2 \sin\theta} (\sqrt{1-f(r)} + 1) \bigg] + B m = 0 \, .
    \end{aligned}
    \end{equation}
The non-trivial solutions are determined by
    \begin{equation}
        \partial_r W_{\pm}(r) = \frac{\bar{B}(r) \pm \sqrt{\bar{\Delta}(r)}}{f(r)},
    \end{equation}
where $\bar{B}(r) = \sqrt{1-f(r)} (\omega + qA_t - a_t) - f(r) a_r$, and $\bar{\Delta}(r) = [\bar{B}(r)]^2 - f(r) H(r)$ with $H(r) = m^2 - (\omega + qA_t - a_t)^2 - 2\sqrt{1-f(r)} (\omega + qA_t - a_t) a_r - \frac{2b_t}{r^2 \sin\theta} (\omega + qA_t - a_t) - f(r) \frac{b_t^2}{r^4 \sin^2 \theta} + f(r) a_r^2$, then near the horizon $f(r) \approx f^\prime(r_H) (r-r_H)$, by the integration around the simple pole $r=r_H$, we find
    \begin{equation}
        W_{+} = \frac{2i\pi(\omega + qA_t(r_H) - a_t(r_H))}{f^\prime(r_H)} \quad \text{and} \quad W_{-} =0 \, .
    \end{equation}
For antifermions $a_t(r_H) \to -a_t(r_H)$, then the Hawking temperature with corrections is
    \begin{equation}
        T_H = \frac{f^\prime (r_H)}{4\pi} \begin{cases}
            \frac{E}{E-a_t(r_H)} & \text{for fermions} \\
            \frac{E}{E+a_t(r_H)} & \text{for antifermions}
        \end{cases} \, .
        \label{MHT-fermions-SME}
    \end{equation}
Here in general, $E=\omega + qA_t(r_H)$, with the change $q \to -q$ for antifermions.

For gravitinos~\eqref{LIV-SME-32}, we use the WKB anzats \eqref{gravitino-ansatz}, again we use the generalized gamma matrices \eqref{Gamma-P}, with the expansion for spinor coefficients and semiclassical action, $I$, with the HJ ansatz \eqref{HJ-ansatz-gravitino} in $t-r$ regime we obtain two systems of equations,
    \begin{equation}
    \begin{aligned}
        d_{\nu} &\bigg[ -(\omega - a_t) - (1 + \sqrt{1-f(r)}) (\partial_r W + a_r) \\
        &\quad - (1 + \sqrt{1-f(r)}) \frac{b_t}{r^2 \sin\theta}\bigg] + b_{\nu} m =0 \, , \\
        b_{\nu} &\bigg[ -(\omega - a_t) + (1- \sqrt{1-f(r)})(\partial_r W + a_r) \\
        &\quad - (1- \sqrt{1-f(r)}) \frac{b_t}{r^2 \sin\theta}\bigg] + d_{\nu} m = 0 \, .
    \end{aligned}
    \end{equation}
The other system is equivalent, $b_{\nu} \to c_{\nu}$ and $d_{\nu} \to a_{\nu}$. And for no trivial solutions
    \begin{equation}
        \partial_r W_{\pm}(r)  = \frac{\bar{B}_g(r) \pm \sqrt{\bar{\Delta}_g(r)}}{f(r)} \, ,
    \end{equation}
where $\bar{B}_g(r) = \sqrt{1-f(r)}(\omega -a_t) - f(r) a_r$, $\bar{\Delta}_g(r) = [\bar{B}(r)]^2 - f(r)\bar{H} (r)$, with $\bar{H}(r) = m^2 -(\omega - a_t)^2 -  2\sqrt{1-f(r)}(\omega -a_t) a_r - 2(\omega -a_t)  \frac{b_t}{r^2\sin\theta} - f(r) \frac{b_t^2}{r^4\sin^2\theta} + f(r) a_r^2$. Using $f(r) \approx f^\prime(r_H) (r-r_H)$, near the horizon we obtain $W_{+} = \frac{2i\pi(\omega-a_t)}{f^\prime(r)}$ and $W_{-} = 0$, from $a_t \to -a_t$ for antigravitinos, for $E=\omega$, we obtain
    \begin{equation}
        T_H = \frac{f^\prime (r_H)}{4\pi} \begin{cases}
            \frac{E}{E-a_t(r_H)} & \text{for gravitinos} \\
            \frac{E}{E+a_t(r_H)} & \text{for antigravitinos}
        \end{cases} \, .
        \label{MHT-gravitinos-SME}
    \end{equation}
For $\Lambda \neq 0$, the Reissner-Nordström-dS spacetime is not asymptotically flat and therefore does not possess an asymptotic region where an observer at infinity can be defined. Instead, the static region accessible to observers lies between the black-hole and cosmological horizons~\cite{Lake1979reissner}. In asymptotically AdS spacetimes, by contrast, one can define a timelike conformal boundary, commonly referred to as ``AdS infinity,'' relative to which the conserved energy can be defined. Nevertheless, both spacetimes admit the Killing vector $\partial_t$, and $\omega$ represents the conserved Killing energy of an emitted particle.

One can show that once we choose a reference frame to define the background vectors $a_\mu^f$, the modified Hawking temperature has the same form as in the dR case.
\section{Thermodynamics}
\label{Termo}
We will derive the thermodynamics assuming only the matter sector is affected and discuss the challenges of generalization.
According to the first law of thermodynamics for the RN-BH 
    \begin{equation}
        dM = T_0 dS + A_t(r_H) dQ \, .
        \label{First-law}
    \end{equation}
where $T_0 = \sqrt{M^2 -Q^2}/[2\pi r_H^2]$,  $r_H^2 = 2M^2 + 2M\sqrt{M^2 - Q^2} - Q^2$ with $r_H=r_H(M,Q)$. 

Since the quantum corrections determined in the previous section yield $T_H=T_0 h(E(\omega,q, Q, M))$, where $h(E(\omega,q, Q, M))$ is a function associated with the correction that depends on the emitted particle energy, then Eq.~\eqref{First-law} is no longer valid, as it breaks the classical thermodynamic structure. Instead, a generalized quantum formulation is adopted as follows: From energy and charge conservation, the total Arnowitt--Deser--Misner (ADM) mass and charge of the BH-particle system remain fixed, while the mass and charge of the BH itself may fluctuate. In the presence of LIV terms, the quantity $\omega=-\partial_t I$ remains well defined, provided that $I$ is a solution of the modified HJ equation, since the LIV terms affect only the matter sector.
When a particle is emitted from the BH, its mass and charge reduce to $M-\omega$ and $Q-q$, respectively. This generalizes the so-called self-gravitation~\cite{Kraus1995self, Parikh2004energy}, i.e., we consider the back-reaction due to the energy and charge conservation; the imaginary part of the true action is
    \begin{equation}
        2Im I_0 = \int_\gamma \frac{d}{dE} \left(\frac{E}{T_H}  \right) dE \, ,
        \label{Im-non-self}
    \end{equation}
where $\gamma$ is the path of integration from $(0,0) \to (\omega,q)$. By changing $M \to u =M-\omega^\prime$ and $Q \to v=Q-q^\prime$ in \eqref{Im-non-self} for RN-BH at the order $L^3$ for dR, we have

\begin{strip}
    \begin{equation}
    \begin{aligned}
        2Im I^{dR}
        &= -2\pi \bigg\{ (1+mL) \int_{(M,Q)}^{(M-\omega,Q-q)} \frac{ [u + \sqrt{u^2 -v^2}]^2}{\sqrt{u^2 - v^2}} \left[du - \frac{v}{[u + \sqrt{u^2 -v^2}]}dv \right] \\
        &\quad + 3(1+mL)L^2 \int_{(M,Q)}^{(M-\omega,Q-q)} \frac{ [u + \sqrt{u^2 -v^2}]^2}{\sqrt{u^2 - v^2}} \left(\omega^\prime - \frac{v}{[u + \sqrt{u^2 -v^2}]}q^\prime \right)^2 \left[du - \frac{v}{[u + \sqrt{u^2 -v^2}]}dv \right] \bigg\} \, .
    \end{aligned} \label{Im-L3}
    \end{equation}
\end{strip}

We can make the integrals choose the line from $(0,0) \to (\omega,q)$: $\omega^\prime(s) = s\omega$ and $q^\prime(s) = sq$ with $s \in [0,1]$, then $u(s)=M-\omega s$ and $v(s)=Q - q s$, so $du = - \omega ds$ and $dv=-qds$, 
    \begin{equation}
        2Im I^{dR}
        =-2\pi (1+mL)\left[ F_0 + 3L^2 F_1\right] \, ,
        \label{Im-dR-p}
    \end{equation}
where $F_0= F_0 (M,\omega,Q,q)$ and $F_1= F_1 (M,\omega,Q,q)$ are give by
    \begin{align}
        F_0 &= - \int_{0}^1 \frac{ [R(s)]^2}{\sqrt{u(s)^2 - v(s)^2}} \left[\omega - \frac{q v(s)}{[R(s)]} \right] ds \, , \label{F_0}\\
        F_1 &= - \int_{0}^1 \frac{ [R(s)]^2}{\sqrt{u(s)^2 - v(s)^2}} \left[\omega - \frac{q v(s)}{[R(s)]} \right]^3 s^2 ds \, , \label{F_1}
    \end{align}
where $R(s)=u(s) + \sqrt{u(s)^2 -v(s)^2}$ is the BH radius, which changes during particle emission.

For the SME corrections, $\frac{E}{T_H} = \frac{E \pm a_t(r_H)}{T_0}$. To incorporate this correction, we introduce the auxiliary quantity $E^\prime = E \pm a_t(r_H)$. Recall that $E$ is the conserved quantity associated with the time-like Killing vector. Then, Eq.~\eqref{Im-non-self} becomes
    \begin{equation}
        2Im I^{SME} = \int_{\gamma^\prime} \frac{d}{dE^\prime} \left(\frac{E^\prime}{T_0}  \right) dE^\prime \, ,
    \end{equation}
where $\gamma^\prime$ is the integration path, from $(0,0) \to (\bar{\omega},q)$, with $\bar{\omega} = \omega \pm a_t(r_H)$. And then using the same parametrization for $u(s)$ and $v(s)$, with the change $\omega \to \bar{\omega}$, we obtain
    \begin{equation}
        2Im I^{SME}
        =-2\pi F_0(M,\bar{\omega},Q,q) \, .
    \end{equation}
The integral Eq.~\eqref{F_0} is
    \begin{equation}
        F_0 = \frac{1}{2} \begin{cases}
            [r_H^2(M-\omega,Q-q) - r_H^2(M,Q)] &\text{for dR} \\
            [r_H^2(M-\bar{\omega},Q-q) - r_H^2(M,Q)] & \text{for SME}
        \end{cases}
    \end{equation}
Notice that $ 2\pi F_0$ matches the result obtained from the Bekenstein--Hawking entropy formula
$S = \frac{A}{4}$,
where $A = 4\pi r_H^2$ is the BH area, in agreement with the first law~\eqref{First-law} ($a_t(r_H)=0$ in the SME). In the SME case, the entropy change is (see the Supplementary Material for details)
    \begin{equation}
        \Delta S_{B-H}^{SME} = \pi \left[r_H^2(M- \omega \mp a_t(r_H) ,Q-q) - r_H^2(M,Q) \right] \, ,
        \label{DS-dR}
    \end{equation}
where $+$($-$) sign is for fermions (antifermions); recall that for gravitinos/antigravitinos $q=0$. The complete expression for $F_1$ (see Eq.~\eqref{F_1}) is quite lengthy. For simplicity, we therefore restrict our attention to the Schwarzschild BH case, where an expansion up to order $L^3$ is unnecessary (see Eq.~\eqref{Im-L3}) and the change of Bekenstein--Hawking entropy is
    \begin{equation} 
        \Delta S_{B-H}^{dR} = -8\pi \omega (1+mL) \left[ \frac{M-\omega}{1-\omega^2 L^2} - \frac{\ln (1-\omega^2 L^2)}{2\omega L^2}\right]\, .
        \label{DS-SME}
    \end{equation}
In the previous results, $\Gamma \sim \exp (\Delta S_{B-H})$, where $\Delta S_{B-H}$ is the change in the Bekenstein--Hawking entropy from the viewpoint of Parikh and Wilczek. In fact, at first order in $\omega$, $q$, and $\omega^3L^2$ for dR, we recover the Hawking temperatures $T_H$ for the corresponding cases (see Section~\ref{QT}).

The methods described above apply to static black holes with a fixed cosmological constant, although the treatment becomes more involved and requires careful attention to subtleties.  
In asymptotically AdS spacetimes, mass is defined as the conserved charge associated with a time-translation Killing field (see~\cite{Ashtekar2000asymptotically}). In contrast, for asymptotically dS spacetimes, the absence of spatial infinity makes it challenging to define global ADM-like quantities.  
In such cases, conservation laws apply locally or at horizon boundaries, so any exchange of mass or charge between the black hole and a particle must preserve the total charge within both the event horizon and the cosmological horizon (see, for example,~\cite{Dolan2019definition}).

\section{Discussion}
\label{Discussion}
Notice that changing $L \to -L$ does not alter the fact that the Hawking temperature decreases in the results for deformed relativity in Eq.~\eqref{MHT-fermions} and Eq.~\eqref{MHT-gravitinos-L} for $LE^2 >m$, or in the massless case. Remarkably, the result obtained in Eq.~\eqref{MHT-fermions} is not consistent with those from~\cite{Li2017application, Li2021lorentz}, since the gamma matrices representation used in it does not reproduce the MDR in Eq.~\eqref{LIV-DR-fermions}. Moreover, when working in BL coordinates, the static observer is described by the choice of gamma matrices~\eqref{Gamma-Sing}. In this case, we find
    \begin{equation}
        \partial_r W_{\pm}(r) = \pm \sqrt{\frac{(\omega + q A_t)^2 - f(r) m^2}{f(r)^2\left(1 - \frac{L^2 (\omega + q A_t)^2}{f(r)}\right)}} \, ,
        \label{W-ill}
    \end{equation}
which suggests that one might expand the denominator under the assumption that $\left|\frac{L^2 (\omega + q A_t)^2}{f(r)}\right| \ll 1$. However, this approximation becomes invalid arbitrarily close to the event horizon, where $f(r_H) = 0$. Fortunately, can avoid the pathological behavior of BL coordinates by employing regular coordinate systems, as we have done in this work. Another interesting example arises in the study of the generalized uncertainty principle, where the effect: $p_0 \to p_0$ and $p_k \to [1 + \beta (p_0^2 - {\bf p}^2)] p_k $ is proposed~\cite{Kober2010gauge}. In this case, the dispersion relation in flat spacetime suggests that the energy is lower than that predicted by Lorentz invariance,
$(p_0)^2 = m^2 + {\bf p}^2 - \frac{\beta m^2 ((p_0)^2 -{\bf p}^2)}{(1 + \beta ((p_0)^2 - {\bf p}^2)]}$, and therefore, the Hawking temperature decreases, as shown in~\cite{Chen2014effects}.

On the other hand, working within the simplified case of the SME, ${\bf b}=0$, we find that the nonzero component $b_t$ of the background vector $b_\mu$, associated with a chiral current that distinguishes between spin projections, does not lead to modifications in the Hawking temperature, nor does the radial component of the background vector, $a_\mu$. Thus, the only component that directly affects the temperature is $a_t(r_H) \propto a_0$, which allows us to distinguish between fermions and antifermions through the transformation properties of its associated current under charge conjugation. Returning to the case of deformed relativity, we can notice that changing $L \to -L$ is not a true description for fermions or antifermions as in the SME case, since in the massless case this does not produce any change, and in fact the dispersion relation is valid in the ultra-relativistic regime~\cite{Ellis2004synchrotron}, which shows an ambiguity that can be removed if we impose that the true equation of motion is Eq.~\eqref{LIV-DR-fermions}. See the supplementary material for the mathematical details of the calculations of previous assertions.

Notice that the only requirement to incorporate the SME in Hawking is to have a finite value for $a_t(r_H)$, and this works like a chemical potential that leads to asymmetric Hawking radiation between fermions and antifermions. A particularly relevant example in this context is that of primordial black holes (PBHs), which can be described by the Schwarzschild metric and are believed to have formed from high-density fluctuations in the early universe~\cite{Carr1974black}. The temperatures derived in Eq.~\eqref{MHT-fermions-SME} and Eq.~\eqref{MHT-gravitinos-SME} may leave observable imprints in the evaporation spectrum of PBHs, potentially suggesting a mechanism for baryon asymmetry in the early universe. The important feature of this mechanism is whether explicit Lorentz-symmetry breaking can give rise to a nonvanishing and finite $a_t(r_H)$~\cite{Kostelecky2021backgrounds}, or whether spontaneous Lorentz-symmetry breaking, as realized in Bumblebee gravity models, can likewise yield a finite $a_t(r_H)$. In the latter case, the Lorentz-violating background modifies the spacetime geometry (see, for example,~\cite{Xu2023static, Mai2023extended, Zhu2026full, Zhu2026new}), so the standard thermodynamic relations of Einstein gravity may no longer apply in their usual form~\cite{An2024notes}. Moreover, obtaining a nonzero LIV term $a_\mu^f$ is crucial for identifying a realistic model~\cite{Kostelecky2011matter}. A detailed investigation of these issues is left for future work.

\section{Summary and conclusions}
\label{Summary}
We have presented a comprehensive treatment of corrections to the Hawking effect derived from QG theories that produce LIV, showing that, given an MDR, the qualitative impact on the HR can be anticipated if the changes are uniquely determined, but explicit corrections are nontrivial. This means that if the MDR includes terms that increase (decrease) the energy relative to that predicted by Lorentz invariance, this leads to an acceleration (deceleration) of BH evaporation.

Our results demonstrate that SME-induced modifications of Hawking radiation can generate an asymmetry in the production of fermions and antifermions, suggesting a novel avenue through which PBHs may contribute to the generation of matter-antimatter asymmetries in the early Universe.

The influence of back-reaction effects on BH thermodynamics has been extensively studied. Such effects modify both the entropy and the Hawking temperature through the conservation of energy and charge during the emission process. For charged black holes, the resulting correction depends on the evolution of both the mass and charge parameters and may either enhance or suppress the effective Hawking temperature. Understanding how these back-reaction effects interplay with corrections from LIV remains an interesting open problem that deserves further investigation.
\section*{Acknowledgements}
I.P.C. thanks SECIHTI for the scholarship as ``Ayudante de Investigador SNII III'' and H. Aguilera-Trujillo thanks SECIHTI (formerly CONACYT) for the scholarship no. 809437. 



\bibliographystyle{unsrt} 
\bibliography{Refs}





\onecolumn

\setcounter{equation}{0}
\renewcommand{\theequation}{S\arabic{equation}}

\begin{center}
    \textbf{\large On modified Hawking radiation under quantum gravity effects in the fermion sector} \\ 
    \vspace{0.5cm}
    { \it \large Supplementary material}\\ 
    \vspace{0.5cm}
    {Ivan P-Castro, Héctor Aguilera-Trujillo, Hugo García-Compeán, Abdel Pérez-Lorenzana}
\end{center}
\subsection*{Chiral Basis and $t-r$ regime}
\label{Spinors-chiral}
Measuring spin in the radial direction, i.e., the direction of the accelerating observer, the spinors take the form in the usual chiral basis given in Eq.~\eqref{Gamma-chiral} in the main text,
    \begin{equation}
        u_{\uparrow}({\bf p}) =
        \begin{pmatrix}
        \sqrt{E-p_{r}} \\
        0 \\
        \sqrt{E + p_r} \\
        0
        \end{pmatrix} \, , \quad
        u_{\downarrow}({\bf p}) =
        \begin{pmatrix}
        0 \\
        \sqrt{E+p_{r}} \\
        0 \\
        \sqrt{E-p_{r}}
        \end{pmatrix} \, , \quad 
        v_{\uparrow}({\bf p}) =
        \begin{pmatrix}
        \sqrt{E-p_{r}} \\
        0 \\
        -\sqrt{E+p_{r}} \\
        0
        \end{pmatrix} \, , \quad
        v_{\downarrow}({\bf p})=
        \begin{pmatrix}
        0 \\
        -\sqrt{E+p_{r}} \\
        0 \\
        \sqrt{E-p_{r}}
        \end{pmatrix} \,.
   \label{Chiral-spinors-radial}
    \end{equation}
It is notable that on this basis, the limit $m \to 0$ recovers the helicity states, and similarly, for gravitinos in flat spacetime, using the Clebsch-Gordan coefficients, we can construct the vector spinors:
    \begin{equation}
    \begin{aligned}
        u^\mu_{3/2} ({\bf k}) &= \epsilon_{+}({\bf k}) u_{\uparrow}({\bf p}) \, , \quad
        u^\mu_{1/2} ({\bf k})
        = \sqrt{\frac{2}{3}} \epsilon_0({\bf k}) u_{\uparrow}({\bf p}) + \frac{1}{\sqrt{3}} \epsilon_{+}({\bf k}) u_{\downarrow}({\bf p}) \, , \\
        u^\mu_{-3/2} ({\bf k}) &= \epsilon_{-}({\bf k}) u_{\downarrow}({\bf p}) \, , \quad
        u^\mu_{-1/2} ({\bf k}) = \sqrt{\frac{2}{3}} \epsilon_{0}({\bf k}) u_{\downarrow}({\bf p}) + \frac{1}{\sqrt{3}} \epsilon_{-}({\bf k}) u_{\uparrow}({\bf p}) \, .
    \end{aligned} \label{Gravitinos-FS}
    \end{equation}
Here $\epsilon_0({\bf k}) = \frac{1}{m}(k_r, 0,0, E)$ and $\epsilon_{\pm} ({\bf k}) = \frac{1}{\sqrt{2}} \left( 0, 1,  \pm i,  0\right)$ are the polarization vectors for massive spin-$1$ field.

Instead of Eq.~\eqref{Gamma-chiral}, in the East Coast signature, which obeys the Clifford algebra $\text{Cliff}(1,3)$, we can construct the chiral basis as
    \begin{equation}
        \tilde{\gamma}^a = \begin{pmatrix}
            0 & \sigma^a \\
            -\bar{\sigma}^a & 0 
        \end{pmatrix} \, ,
    \end{equation}
This yields a similar Dirac spinor. However, these representations are not related by unitary transformation $U$, $\tilde{\gamma}_a = U \tilde{\gamma}^\prime_a U^\dagger$, because $\text{Cliff}(1,3) \not\simeq \text{Cliff}(3,1)$. In practice, one obtains the same observables (e.g., the same dispersion relations) provided the equations of motion or Lagrangian densities are constructed appropriately.
\subsection*{Modified dispersion relations}
Using \eqref{LIV-DR-fermions}, and the solution $\psi(x) = \tilde{\psi}({\bf p}) e^{-ip_a x^a}$, we obtain
     \begin{equation}
        \left(\tilde{\gamma}^a p_a - m + L \tilde{\gamma}^0 p_0 \tilde{\gamma}^j p_j \right) \tilde{\psi}(p) = 0 \, .
    \end{equation}
From \eqref{Gamma-chiral} and the usual Pauli matrices
    \begin{equation}
        \sigma^1 = \left(\ba{cc} 0 & 1 \\
        1 & 0 \ea\right) \, , \quad \sigma^2 = \left(\ba{cc} 0 & -i \\
        i & 0 \ea\right) \, , \quad \sigma^3 = \left(\ba{cc} 1 & 0 \\
        0 & -1 \ea\right) \, ,
        \label{Pauli-matrices}
    \end{equation}
we obtain
    \begin{equation*}
    \begin{aligned}
        0 &= \det (\slashed{p} + L \tilde{\gamma}^0 \tilde{\gamma}^j p_0 p_j - m) = \left| \ba{cc} - \sigma^j L p_0 p_j - m & \sigma^a p_a \\
        \bar{\sigma}^a p_a & \sigma^j L p_0 p_j - m \ea\right| \\
        &= \left| \ba{cccc} -L p_0 p_3 -m & -L p_0(p_1 - ip_2) & p_0 + p_3 & p_1 -ip_2 \\
        -L p_0(p_1 + ip_2) & L p_0 p_3 - m & p_1 + ip_2 & p_0 - p_3 \\
        p_0 - p_3 & -(p_1 -ip_2) & L p_0 p_3 - m & L p_0(p_1 -ip_2) \\
        -(p_1 + ip_2) & p_0 + p_3 & L p_0(p_1 + ip_2) & -L p_0 p_3 -m
        \ea\right| = [m^2 + {\bf p}^2 - (p_0)^2(1 + L^2 {\bf p}^2) ]^2.
    \end{aligned}
    \end{equation*}
Then the MDR is
    \begin{equation}
        (p_0)^2 = m^2 + {\bf p}^2 - (Lp_0)^2 {\bf p} ^2 ,
    \end{equation}
as \eqref{MDR-DR} with $\eta_{+} =1$ and $n =2$, we can obtain the same result for gravitinos from \eqref{LIV-DR-gravitinos}; $(k_0)^2 = m^2 + {\bf k}^2 - (Lk_0)^2 {\bf k}^2$. 

On the other hand, when we use the East Coast convention, the correct modified Dirac equation is
    \begin{equation}
        (\tilde{\gamma}^a \partial_a + \bar{m} - L \tilde{\gamma}^0 \partial_0 \tilde{\gamma}^j \partial_j) \psi (x) = 0 \Rightarrow (-i\tilde{\gamma}^a p_a + \bar{m} + L \tilde{\gamma}^0 \tilde{\gamma}^j p_0 p_j) \tilde{\psi} ({\bf p}) =0 \, .
    \end{equation}
Here, we have used $\psi(x) = \tilde{\psi}({\bf p}) e^{-ip_ax^a}$, with 
    \begin{equation}
    \begin{aligned}
        0 &= \det[-i \tilde{\gamma}^a p_a + \bar{m} + L  (\tilde{\gamma}^0 p_0)(\tilde{\gamma}^j p_j)] = \left| \ba{cc} L p_0 \sigma^j p_j + \bar{m} & -i(\sigma^0 p_0+ \sigma^j p_j) \\ 
        -i(-\sigma^0 p_0+ \sigma^j p_j) & -L p_0 \sigma^j p_j + \bar{m}
        \ea\right| \\
        &= \left|\ba{cccc} L p_0 p_3 + \bar{m} & L p_0 (p_1 - ip_2) & -i(p_0 + p_3) & -i(p_1 - ip_2) \\
        L p_0(p_1 + ip_2)  & -L p_0 p_3 + \bar{m} & -i(p_1 + ip_2) & -i(p_0 - p_3) \\
        -i(-p_0 + p_3) & -i(p_1 - ip_2) & -L p_0 p_3 + \bar{m} & -L p_0 (p_1 - ip_2) \\
        -i(p_1 + ip_2) & -i(-p_0 - p_3) & -L p_0(p_1 + ip_2)  & L p_0 p_3 + \bar{m}
        \ea\right| = [\bar{m}^2 + {\bf p}^2 - (p_0)^2 (1 + L^2 {\bf p}^2)]^2.
    \end{aligned}
    \end{equation}
This MDR is the same as that obtained for spin-$\frac{1}{2}$ fermions and gravitinos. We can notice two relevant things: (i) The dispersion relation does not depend on the sign convention used in the metric, nor on the plane wave solution we choose, (ii) Changing the parameter $L \to -L$ gives the same dispersion relation, and this is below that predicted by Lorentz invariance; in fact, it is possible to choose $L \to iL$, and that is actually the case that has been presented in the works~\cite{Li2017application,Li2021lorentz}.

Now, in SME for $k=0$. Following the previous prescription, but now for gravitinos
    \begin{equation*}
    \begin{aligned}
        0 &= \det (\slashed{k}-\slashed{a} - \gamma_5 \slashed{b} - m) = \left| \ba{cc}- m & \sigma^c (k_c -a_c + b_c) \\
        \bar{\sigma}^c (k_c - a_c - b_c) & - m \ea\right| \quad \text{If $q_c = k_c - q_c + b_c, s_c = k_c - a_c - b_c$} \\
        &= \left| \ba{cccc} -m & 0 & q_0 + q_3 & q_1 -iq_2 \\
        0 & - m & q_1 + iq_2 & q_0 - q_3 \\
        s_0 - s_3 & -(s_1 -is_2) & - m & 0 \\
        -(s_1 + is_2) & s_0 + s_3 & 0 & -m
        \ea\right| \\
        &=  -m^2 (s_0 + s_3)(q_0 - q_3) + m^2 [m^2 + (s_1 - is_2)(q_1 + iq_2)] - m^2 (q_0 + q_3)(s_0 - s_3) + (q_0^2 - q_3^2) [(s_0^2 - s_3^2) - (s_1^2 + s_2^2)] \\
        &\quad + m^2 (q_1 -i q_2) (s_1 + is_2) - (q_1^2 + q_2^2) [(s_0^2 - s_3^2) - (s_1^2 + s_2^2)] \\
        &= [(k-a)^2 -b^2 - m^2]^2 + 4b^2 (k-a)^2 - 4 [b^c (k_c - a_c)]^2.
    \end{aligned}
    \end{equation*}
Here $(k-a)^2 = (k^c-a^c) (k_c-a_c)$ and $b^2 = b^c b_c$. In general, we need to solve a fourth-degree equation in $p_0$ to determine the energy spectrum. But in the simples case with ${\bf b} =0$, we obtain four solutions    
    \begin{equation}
        k_0 = [m^2 + \left( |{\bf p} - {\bf a}| + (-1)^\alpha b_0 \right)^2]^{1/2} + a_0 \quad \text{and} \quad k_0 = -[m^2 + \left( |{\bf p} - {\bf a}| + (-1)^\alpha b_0 \right)^2]^{1/2} + a_0
    \end{equation}
with $\alpha = 1,2$. For a proper physical interpretation, only the first two roots are used, which define energies analogous to Eq.~\eqref{f-E}. Recall that to describe antigravitinos we use $a_c \to -a_c$. Finally, an analogous calculation gives us the spectrum for spin-$\frac{1}{2}$ fermions.

Another interesting case is to consider the generalized uncertainty principle (GUP) on the West Coast, where
    \begin{equation}
        p_0 = p_0 \quad \text{and} \quad p_j \to (1+\beta p^2) p_j
    \end{equation}
Then, the modified Dirac equation is
    \begin{equation}
        [i\tilde{\gamma}^a (1-\beta \partial^b \partial_b)\partial_a -m]\psi = 0 \, .
    \end{equation}
So the Dispersion relation is determined by $[\bar{m}^2 + {\bf p}^2 (1 + \beta p^2) - (p_0)^2 (1 + \beta p^2) ]^2 = 0$, and we obtain
    \begin{equation}
        (p_0)^2 = \frac{m^2 + {\bf p}^2 (1 + \beta p^2)}{(1 + \beta p^2)} = m^2 + {\bf p}^2 - \frac{\beta m^2 ((p_0)^2 -{\bf p}^2)}{(1 + \beta ((p_0)^2 -{\bf p}^2)]} \, .
    \end{equation}
Therefore, this dispersion relation gives us an energy of a free massive fermion slightly lower than that established by Lorentz invariance. If we use the East Coast convention, we simply need to change $p_0 = p_0 \quad \text{and} \quad p_j \to (1-\beta p^2) p_j$, and with the equation of motion, $[\tilde{\gamma}^a (1+\beta \partial^b \partial_b)\partial_a + \bar{m}]\psi = 0$, we will get the same result.
\subsection*{Hawking temperatures}
For dR, we can write the Dirac-like equation~\eqref{LIV-DR-12} for the spin-up case,
    \begin{equation}
    \begin{aligned}
        0 &=i\gamma^\mu \left(\partial_\mu -i\frac{q}{\hslash}A_\mu + \Omega_\mu \right) \psi_{\uparrow} - \frac{m}{\hslash} \psi_{\uparrow} - \hslash L \gamma^t \gamma^j \left(\partial_t -i\frac{q}{\hslash}A_t + \Omega_t \right) \left(\partial_j + \Omega_j \right) \psi_{\uparrow} \, .
        \label{spin-up-DR}
    \end{aligned}
    \end{equation}
Using the Painlevé-Gullstrand coordinates for fermions with a freely falling observer~\eqref{Gamma-P}, then from Eq.~\eqref{spin-up-DR}, we obtain 
    \begin{equation}
    \begin{aligned}
        B \left[ (\partial_t I_{\uparrow} - qA_t) + \left( 1 - \sqrt{1-f(r)} \right) (\partial_r I_{\uparrow}) \right] 
        &= - A \bigg[ m + \left( \sqrt{1-f(r)} + 1 \right) L (\partial_t I_{\uparrow} -qA_t)(\partial_r I_{\uparrow})\bigg] \, , \\
        A \left[ (\partial_t I_{\uparrow} - q A_t) - \left( 1 + \sqrt{1-f(r)} \right) (\partial_r I_{\uparrow}) \right] 
        &= -B \bigg[ m + \left( \sqrt{1-f(r)} - 1 \right) L (\partial_t I_{\uparrow} -qA_t) (\partial_r I_{\uparrow})\bigg] \, ,
    \end{aligned}
    \end{equation}
and the angular part becomes
    \begin{equation}
    \begin{aligned}
        \frac{1}{r} \left[B + A L (\partial_t I_{\uparrow}-qA_t) \right] \left[ (\partial_\theta I_{\uparrow}) + \frac{i}{\sin\theta} (\partial_\varphi I_{\uparrow})\right]  &= 0, \\
        - \frac{1}{r} \left[A + B L (\partial_t I_{\uparrow} -q A_t) \right] \left[ (\partial_\theta I_{\uparrow}) + \frac{i}{\sin\theta} (\partial_\varphi I_{\uparrow})\right] &= 0.
    \end{aligned}
    \end{equation}
From the HJ anzats given in Eq.~\eqref{HJ-spin-up}, we can impose $(\partial_\theta I_{\uparrow}) + \frac{i}{\sin\theta} (\partial_\varphi I_{\uparrow}) = (\partial_\theta J) + \frac{i}{\sin\theta} (\partial_\varphi J) =0$, and integrating around the pole $r=r_H$ the solution given in Eq.~\eqref{W-spin-up}, we get
    \begin{equation}
        W_{\pm} = \int_{r_H-\epsilon}^{r_H + \epsilon} \frac{B(r) \pm \sqrt{\Delta(r)}}{A(r)} dr = \frac{B(r_H) \pm B(r_H)}{f^\prime (r_H) (1-E^2 L^2)} \int_{r_H-\epsilon}^{r_H + \epsilon} \frac{dr}{r-r_H} = \begin{cases}
            \frac{2i\pi E (1 + mL)}{f^\prime(r_H) (1 - E^2 L^2)} & \text{for $+$} \\
            0 & \text{for $-$}
        \end{cases} \, .
    \end{equation}
Then, the rate for quantum tunneling is
    \begin{equation}
        \Gamma = \frac{P[out]}{P[in]} = \exp\{-2 Im I_{0,\uparrow}\} = \exp\{-2Im W_{+}\} = \exp\left\{ - \frac{4\pi E (1 + mL)}{f^\prime(r_H) (1 - E^2 L^2)} \right\} = \exp\left\{ - \frac{E}{T_H} \right\} \, .
        \label{rate-qt}
    \end{equation}
For a static observer, we obtain the representation for gamma matrices~\eqref{Gamma-GP}. In this case, we need to solve the following system of equations for $t-r$ variables
    \begin{equation}
    \begin{aligned}
        B \left[ \frac{1}{\sqrt{f(r)}} \left(1 + \sqrt{1-f(r)} \right) (\partial_t I_{\uparrow} - q A_t) + \sqrt{f(r)} (\partial_r I_{\uparrow}) \right] + A \left[ m + \left(1 + \sqrt{1-f(r)} \right) L (\partial_t I_{\uparrow}-qA_t)(\partial_r I_{\uparrow}) \right] &= 0 \, ,\\
        A \left[ \frac{1}{\sqrt{f(r)}} \left(1 - \sqrt{1-f(r)} \right) (\partial_t I_{\uparrow} - q A_t) - \sqrt{f(r)} (\partial_r I_{\uparrow}) \right] + B \left[ m - \left(1 - \sqrt{1-f(r)} \right) L (\partial_t I_{\uparrow}-qA_t)(\partial_r I_{\uparrow}) \right] &= 0 \, .
    \end{aligned}
    \end{equation}
And this leads to the same rate obtained in~\eqref{rate-qt}.
    
Similarly, for gravitinos we consider
    \begin{equation}
        0 = i\gamma^\mu \left(\partial_\mu + \Omega_\mu \right) \Psi_\nu - i\gamma^\mu \Gamma_{\mu\nu}^\lambda \Psi_\lambda - \frac{m}{\hslash} \Psi_\nu - \hslash L \gamma^t \gamma^j \left(\partial_t + \Omega_t \right) \left(\partial_j + \Omega_j \right) \Psi_\nu \, .
        \label{gravitino-DR}
    \end{equation}
From \eqref{gravitino-DR}
and \eqref{gravitino-ansatz}, in $t-r$ regime, we obtain
    \begin{equation}
    \begin{aligned}
        c_{\nu} \left[\frac{1}{\sqrt{f(r)}} \left(1 + \sqrt{1-f(r)} \right) (\partial_t I)+ \sqrt{f(r)} (\partial_r I) \right] + a_{\nu} \left[ m + \left( 1 + \sqrt{1-f(r)} \right) L(\partial_t I \partial_r I) \right] &= 0 \, ,\\
        a_{\nu} \left[ \frac{1}{\sqrt{f(r)}} \left(1 - \sqrt{1-f(r)} \right) (\partial_t I) - \sqrt{f(r)} (\partial_r I_0)\right] + c_{\nu} \left[ m - \left( 1- \sqrt{1-f(r)} \right) L(\partial_t I \partial_r I) \right] &= 0 \, ,
    \end{aligned}
    \end{equation}
and 
    \begin{equation}
    \begin{aligned}
        d_{\nu} \left[ \frac{1}{\sqrt{f(r)}} \left(1 - \sqrt{1-f(r)} \right) (\partial_t I) - \sqrt{f(r)} (\partial_r I)\right] + b_{\nu} \left[ m - \left( 1- \sqrt{1-f(r)} \right) L(\partial_t I \partial_r I) \right] &= 0 \, ,\\
        b_{\nu} \left[\frac{1}{\sqrt{f(r)}} \left(1 + \sqrt{1-f(r)} \right) (\partial_t I)+ \sqrt{f(r)} (\partial_r I) \right] + d_{\nu} \left[ m + \left( 1 + \sqrt{1-f(r)} \right) L(\partial_t I \partial_r I) \right] &= 0 \, .
    \end{aligned}
    \end{equation}
Notice that the systems are equivalent $c_{\nu} \to b_{\nu}$ and $a_{\nu} \to d_{\nu}$, and we obtain the nontrivial solutions after using HJ ansatz
    \begin{equation}
        \partial_r W_{\pm}(r) = \frac{B_g(r) \pm \sqrt{\Delta_g(r)}}{A_g(r)} \, ,
        \label{W-gravitino}
    \end{equation}
where $A_g(r) =  f(r) (1 - \omega^2 L^2)$ , $
C_g(r)= m^2 - \omega^2$ , $
B_g(r) = \sqrt{1-f(r)} \omega(1 + mL) $ and $\Delta_g(r) = [B_g(r)]^2 - A_g(r) C_g(r)$. This leads
    \begin{equation}
        W_{\pm} = \int_{r_H-\epsilon}^{r_H + \epsilon} \frac{B_g(r) \pm \sqrt{\Delta_g(r)}}{A_g(r)} dr = \frac{B(r_H) \pm B(r_H)}{f^\prime (r_H) (1-E^2 L^2)} \int_{r_H-\epsilon}^{r_H + \epsilon} \frac{dr}{r-r_H} = \begin{cases}
            \frac{2i\pi E (1 + mL)}{f^\prime(r_H) (1 - E^2 L^2)} & \text{for $+$} \\
            0 & \text{for $-$}
        \end{cases}
    \end{equation}
But now, $E = \omega$, because there is no electromagnetic interaction, and we obtain the same tunneling rate~\eqref{rate-qt}.

We can analyze 
    \begin{equation}
    \frac{1-E^2L^2}{1 + m(\pm L)}<1 \Leftrightarrow -E^2 L< \pm m \, .
    \end{equation}
This condition is satisfied trivially by $L$, but for $-L$: $\frac{m}{L}<E^2$ is only valid for very high energies.

Using the gamma matrices representation~\eqref{Gamma-Sing}, we use the HJ ansatz $I=-\omega t + W(r) + J(\theta, \varphi) + K$ with $K$ a complex constant; we obtain the system of equations in $t-r$ regime
    \begin{equation}
    \begin{aligned}
        B \left[ -\frac{(\omega + qA_t)}{\sqrt{f(r)}} + \sqrt{g(r)} (\partial_r W) \right] + A \left[ m - L (\omega + q A_t) (\partial_r W) \sqrt{\frac{g(r)}{f(r)}} \right] &= 0 \, , \\
        A \left[ -\frac{(\omega + qA_t)}{\sqrt{f(r)}} - \sqrt{g(r)} (\partial_r W) \right] + B \left[ m + L (\omega +q A_t) (\partial_r W) \sqrt{\frac{g(r)}{f(r)}} \right] &= 0 \, .
    \end{aligned}
    \end{equation}
For $f(r) = g(r)$, we obtain
    \begin{equation}
        (\partial_r W)^2 = \frac{(\omega+qA_t)^2-f(r) m^2}{f(r)(f(r) - L^2 (\omega + q A_t)^2)} = \frac{(\omega+qA_t)^2-f(r) m^2}{f(r)^2( 1 - L^2 (\omega + q A_t)^2/f(r))} \, .
    \end{equation}
That reduces to Eq.~\eqref{W-ill}.

Moreover, if we consider 
    \begin{equation}
        (i \tilde{\gamma}^a \partial_a - m -L\tilde{\gamma}^0 \partial_0 \tilde{\gamma}^l \partial_l) \to ( \tilde{\gamma}^a p_a - m + L\tilde{\gamma}^0 p_0 \tilde{\gamma}^l p_l) \, .
    \end{equation}
we can obtain the dispersion relation from
    \begin{equation*}
    \begin{aligned}
        ( \tilde{\gamma}^b p_b + m + L\tilde{\gamma}^0 p_0 \tilde{\gamma}^k p_k) ( \tilde{\gamma}^a p_a - m + L\tilde{\gamma}^0 p_0 \tilde{\gamma}^l p_l) &= (\tilde{\gamma}^b p_b + m)(\tilde{\gamma}^a p_a - m) + (\tilde{\gamma}^b p_b + m) L\tilde{\gamma}^0 p_0 \tilde{\gamma}^l p_l  + L\tilde{\gamma}^0 p_0 \tilde{\gamma}^k p_k( \tilde{\gamma}^a p_a - m) \\
        &\quad + L^2\tilde{\gamma}^0 p_0 \tilde{\gamma}^k p_k \tilde{\gamma}^0 p_0 \tilde{\gamma}^l p_l \\
        &= p_a^2 - m^2 + L [\tilde{\gamma}^b p_b \tilde{\gamma}^0 p_0 \tilde{\gamma}^l p_l + \tilde{\gamma}^0 p_0 \tilde{\gamma}^k p_k \tilde{\gamma}^a p_a] - L^2\tilde{\gamma}^0 \tilde{\gamma}^0 (p_0)^2 \tilde{\gamma}^k p_k \tilde{\gamma}^l p_l \\
        &= (p_0)^2 - {\bf p}^2 - m^2 + L^2 (p_0)^2 {\bf p}^2
    \end{aligned}
    \end{equation*}
Then, we can obtain a Klein-Gordon-like equation as
    \begin{equation*}
    \begin{aligned}
        0 = (i \gamma^{\mu} {\cal D}_{\mu} + m_f + L\tilde{\gamma}^0 {\cal D}_0 \tilde{\gamma}^l {\cal D}_l )(i \gamma^{\nu} {\cal D}_{\nu} - m_f + L\tilde{\gamma}^0 {\cal D}_0 \tilde{\gamma}^k {\cal D}_k) \psi \approx (i \gamma^{\mu} {\cal D}_{\mu} + m) (i \gamma^{\nu} {\cal D}_{\nu} - m)\psi + L^2\gamma^0 \gamma^l \gamma^0 \gamma^k {\cal D}_0 {\cal D}_l {\cal D}_0  {\cal D}_k \psi \, .
    \end{aligned}
    \end{equation*}
For the static frame defined in BL coordinates, and the usual ansatz~\eqref{12Fermion-ansatz}, after decoupling the angular variables from $t-r$ regime, we find
    \begin{equation}
        \lambda A = \left[ \frac{(\omega + qA_t)^2}{f(r)} -f(r) (\partial_r W)^2 - m_f^2 + L^2 (\partial_r W)^2 (\omega + qA_t)^2 \right] A \, .
    \end{equation}
where $\lambda$ is a separability constant. This solution has the same problem as Eq.~\eqref{W-ill}; therefore, the use of BL coordinates obstructs the extraction of the imaginary part of the semiclassical action.
\subsection{Modified Bekenstein-Hawking entropy}
For charged or non-charged fermions, the modified Hawking temperature under deformed relativity at order $L^3$
    \begin{equation}
        \frac{d}{dE} \left(\frac{E}{T_H} \right) = \frac{d}{dE} \left(\frac{4\pi E}{f^\prime(r_H)} \frac{(1 + mL)}{1-E^2L^2}\right) = \frac{4\pi}{f^\prime(r_H)} \frac{(1+mL)(1+E^2L^2)}{(1-E^2L^2)^2} \approx \frac{4\pi}{f^\prime(r_H)} (1 + mL) (1+3E^2L^2) \, .
    \end{equation} 
In RN case, $\frac{1}{T_0}= \frac{4\pi}{f^\prime(r_H)} =\frac{ 2\pi (M + \sqrt{M^2 -Q^2})^2}{\sqrt{M^2 - Q^2}}$, and we have
    \begin{equation}
    \begin{aligned}
        2 Im I &= \int_{(0,0)}^{(\omega,q)} \frac{1}{T_0}(1+mL) (1+3(\omega^\prime + A_t(r_H)q^\prime)^2L^2) (d\omega^\prime + A_r(r_H)dq^\prime) \\
        &= 2\pi (1+mL)\int_{(0,0)}^{(\omega,q)} \frac{ (M + \sqrt{M^2 -Q^2})^2}{\sqrt{M^2 - Q^2}} (1+3(\omega^\prime + A_t(r_H)q^\prime)^2L^2) (d\omega^\prime + A_r(r_H)dq^\prime) \\
        &\quad + 6\pi (1+mL)L^2 \int_{(0,0)}^{(\omega,q)} \frac{ (M + \sqrt{M^2 -Q^2})^2}{\sqrt{M^2 - Q^2}}(\omega^\prime + A_t(r_H)q^\prime)^2 (d\omega^\prime + A_r(r_H)dq^\prime).
    \end{aligned}
    \end{equation}
Replacing $M \to u = M-\omega^\prime$ and $v=Q-q^\prime$, we obtain
    \begin{equation}
    \begin{aligned}
        2Im I^{dR} &= -2\pi (1+mL) \int_{(M,Q)}^{(M-\omega,Q-q)} \frac{ [(M-\omega^\prime) + \sqrt{(M-\omega^\prime)^2 -(Q-q)^2}]^2}{\sqrt{(M-\omega^\prime)^2 - (Q-q)^2}} \\
        &\quad \times \left[d(M-\omega^\prime) - \frac{Q-q^\prime}{[(M-\omega^\prime) + \sqrt{(M-\omega^\prime)^2 -(Q-q)^2}]}d(Q-q^\prime) \right] \\
        &\quad - 6\pi (1+mL)L^2 \int_{(M,Q)}^{(M-\omega,Q-q)} \frac{ [(M-\omega^\prime) + \sqrt{(M-\omega^\prime)^2 -(Q-q)^2}]^2}{\sqrt{(M-\omega^\prime)^2 - (Q-q)^2}}\\
        &\quad \times \left(\omega^\prime - \frac{Q-q^\prime}{[(M-\omega^\prime) + \sqrt{(M-\omega^\prime)^2 -(Q-q)^2}]}q^\prime \right)^2 \\
        &\quad \times [d(M-\omega^\prime) - \frac{Q-q^\prime}{[(M-\omega^\prime) + \sqrt{(M-\omega^\prime)^2 -(Q-q)^2}]}d(Q-q^\prime)] \\
        &= -2\pi \bigg\{ (1+mL) \int_{(M,Q)}^{(M-\omega,Q-q)} \frac{ [u + \sqrt{u^2 -v^2}]^2}{\sqrt{u^2 - v^2}} \left[du - \frac{v}{[u + \sqrt{u^2 -v^2}]}dv \right] \\
        &\quad + 3(1+mL)L^2 \int_{(M,Q)}^{(M-\omega,Q-q)} \frac{ [u + \sqrt{u^2 -v^2}]^2}{\sqrt{u^2 - v^2}} \left(\omega^\prime - \frac{v}{[u + \sqrt{u^2 -v^2}]}q^\prime \right)^2 \left[du - \frac{v}{[u + \sqrt{u^2 -v^2}]}dv \right] \bigg\} \, ,
    \end{aligned}
    \end{equation}
and then, from here, with the parametrization described after \eqref{Im-L3}, we obtain Eq.~\eqref{Im-dR-p}. While for the Schwarzschild black hole, we do not approximate; the exact expression is
    \begin{equation}
    \begin{aligned}
        2Im I^{dR} &= -8\pi (1+mL) \int_{(M,Q)}^{(M-\omega,Q-q)} (M-\omega^\prime) \frac{1+{\omega^\prime}^2 L^2}{(1-{\omega^\prime}^2 L^2)^2} d(M-\omega^\prime) 
        = 8\pi\omega (1 + mL) \int_0^1 u(s) \frac{1+\omega^2 s^2L^2}{(1-\omega^2 s^2L^2)^2} ds \, .
    \end{aligned}
    \end{equation}

On the other hand, for the SME case, in order to restore the correct $T_H$, we can redefine $\bar{\omega} = \omega \pm a_t$, so
    \begin{equation}
        2Im I = \int_{\gamma^\prime} \frac{d}{dE^\prime} \left( \frac{E^\prime}{T_0} \right) dE^\prime = \int_{\gamma^\prime}  \frac{1}{T_0}  dE^\prime\, ,
    \end{equation}
where $E^\prime = \bar{\omega} + qA_t(r_H)$, and $\gamma^\prime$ is the path from $(0,0) \to (\bar{\omega},q)$. Again we replace $M \to u=M-\omega^\prime$ and $v=Q-q^\prime$
    \begin{equation*}
    \begin{aligned}
        2Im I &= -2\pi \int_{(M,Q)}^{(M-\bar{\omega},Q-q)} \frac{ [(M-\omega^\prime) + \sqrt{(M-\omega^\prime)^2 -(Q-q)^2}]^2}{\sqrt{(M-\omega^\prime)^2 - (Q-q)^2}} \left[d(M-\omega^\prime) - \frac{Q-q^\prime}{[(M-\omega^\prime) + \sqrt{(M-\omega^\prime)^2 -(Q-q)^2}]}d(Q-q^\prime) \right] \\
        &= -2\pi \bigg\{ \int_{(M,Q)}^{(M-\bar{\omega},Q-q)} \frac{ [u + \sqrt{u^2 -v^2}]^2}{\sqrt{u^2 - v^2}} \left[du - \frac{v}{[u + \sqrt{u^2 -v^2}]}dv \right] \bigg\}.
    \end{aligned}
    \end{equation*}
So, with the parametrization described in Section~\ref{Termo}, we arrive at the result in Eq.~\eqref{DS-dR}.
\end{document}